\documentclass[sigconf]{acmart}

\usepackage{amsmath,nccmath,tabularx} 
\usepackage{algorithm}
\usepackage{algorithmic}
\usepackage{color}
\usepackage{graphicx}
\graphicspath{{images/}}
\usepackage{wrapfig,lipsum}
\usepackage{tabularx}
\usepackage{graphicx}
\usepackage{lscape}
\usepackage{subcaption}
\usepackage{latexsym}
\usepackage{pifont}
\usepackage{multirow}

\newcommand{\cmark}{\ding{51}}%
\usepackage{enumitem}
\setlist{leftmargin=6mm}
\usepackage[T1]{fontenc}
\usepackage[utf8]{inputenc}
\usepackage[font=small,labelfont=bf]{caption}

\usepackage{booktabs}

\usepackage{xcolor}
\usepackage{xspace}
\usepackage{bm}
\usepackage{amsmath}
\usepackage{placeins}

\usepackage{balance}

\AtBeginDocument{%
	}

\copyrightyear{2024}
\acmYear{2024}
\setcopyright{rightsretained}
\acmConference[SIGIR '24]{Proceedings of the 47th International ACM SIGIR Conference on Research and Development in Information Retrieval}{July 14--18, 2024}{Washington, DC, USA}
\acmBooktitle{Proceedings of the 47th International ACM SIGIR Conference on Research and Development in Information Retrieval (SIGIR '24), July 14--18, 2024, Washington, DC, USA}
\acmDOI{10.1145/3626772.3657813}
\acmISBN{979-8-4007-0431-4/24/07}
\makeatletter
\gdef\@copyrightpermission{
	\begin{minipage}{0.3\columnwidth}
		\href{https://creativecommons.org/licenses/by-nc-sa/4.0/}{\includegraphics[width=0.90\textwidth]{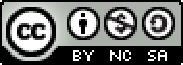}}
	\end{minipage}\hfill
	\begin{minipage}{0.7\columnwidth}
		\href{https://creativecommons.org/licenses/by-nc-sa/4.0/}{This work is licensed under a Creative Commons Attribution-NonCommercial-ShareAlike International 4.0 License.}
	\end{minipage}
	\vspace{5pt}
}
\makeatother

\begin{document}

\title{A Setwise Approach for Effective and Highly Efficient Zero-shot Ranking with Large Language Models}


\author{Shengyao Zhuang}
\affiliation{
	\institution{CSIRO}
	\city{Brisbane}
	\country{Australia}
}
\email{shengyao.zhuang@csiro.au}
\authornote{Corresponding author.}

\author{Honglei Zhuang}
\affiliation{%
	\institution{Google Research}
	\city{Mountain View}
	\country{USA}
}
\email{hlz@google.com}

\author{Bevan Koopman}
\affiliation{%
	\institution{CSIRO}
		\city{Brisbane}
	\country{Australia}
}
\email{bevan.koopman@csiro.au}

\author{Guido Zuccon}
\affiliation{%
	\institution{\mbox{The University of Queensland}}
		\city{St Lucia}
	\country{Australia}
}
\email{g.zuccon@uq.edu.au}

\begin{abstract}
	
	We propose a novel zero-shot document ranking approach based on Large Language Models (LLMs): the Setwise prompting approach.  
	Our approach complements existing prompting approaches for LLM-based zero-shot ranking: Pointwise, Pairwise, and Listwise.
	Through the first-of-its-kind comparative evaluation within a consistent experimental framework and considering factors like model size, token consumption, latency, among others, we show that existing approaches are inherently characterised by trade-offs between effectiveness and efficiency. 
	We find that while Pointwise approaches score high on efficiency, they suffer from poor effectiveness. Conversely, Pairwise approaches demonstrate superior effectiveness but incur high computational overhead.
	Our Setwise approach, instead, reduces the number of LLM inferences and the amount of prompt token consumption during the ranking procedure, compared to previous methods. This significantly improves the efficiency of LLM-based zero-shot ranking, while also retaining high zero-shot ranking effectiveness. We make our code and results publicly available at \url{https://github.com/ielab/llm-rankers}.

\end{abstract}


\maketitle

\section{Introduction}

\begin{figure*}
	\centering
	\includegraphics[width=0.95\linewidth]{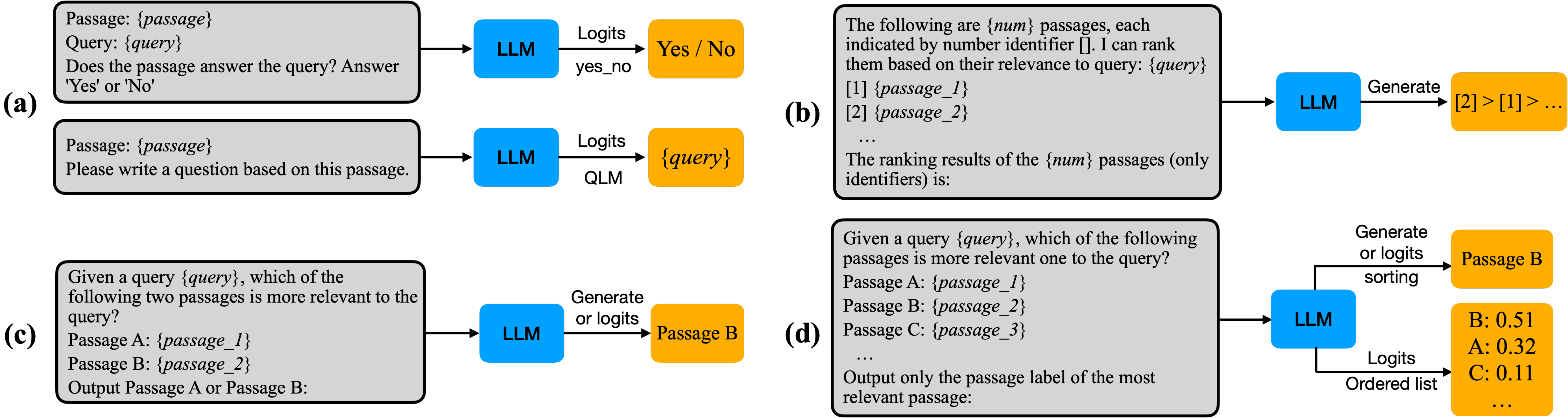}
	\caption{Different prompting strategies. (a) \textit{Pointwise}, (b) \textit{Listwise}, (c) \textit{Pairwise} and (d) our proposed \textit{Setwise.}\vspace{-6pt}}
	\label{fig:methods}
\end{figure*}

Large Language Models (LLMs) such as GPT-3~\cite{brown2020language}, FlanT5~\cite{wei2021finetuned}, Gemini~\cite{team2023gemini}, LLaMa2~\cite{touvron2023llama2}, and PaLM~\cite{chowdhery2022palm} are highly effective across a diverse range of natural language processing tasks under the zero-shot settings~\cite{kojima2022large,brown2020language,agrawal2022large,10.1145/3539618.3591703}.
Notably, these LLMs have also been adapted for zero-shot document ranking.~\cite{liang2022holistic,sachan-etal-2022-improving,sun2023chatgpt,ma2023zero,pradeep2023rankvicuna,qin2023large}.
Using LLMs in zero-shot ranking tasks can be broadly categorized into three main approaches: \textit{Pointwise}~\cite{liang2022holistic,sachan-etal-2022-improving,zhuang-etal-2023-open}, \textit{Listwise}~\cite{sun2023chatgpt,ma2023zero,pradeep2023rankvicuna}, and \textit{Pairwise}~\cite{qin2023large}. These approaches employ different prompting strategies to instruct the LLM to output a relevance estimation for each candidate document and rank documents accordingly. Compared to traditional neural ranking methods~\cite{lin2021pretrained}, these LLM-based rankers do not require any further supervised fine-tuning and exhibit strong zero-shot ranking capabilities.
While these LLM-based zero-shot ranking approaches have been successful individually, it is worth noting that there has been a lack of fair comparison in the literature regarding their effectiveness, and in particular, their efficiency within the exact same experimental framework. This includes factors such as utilizing the same size LLM, evaluation benchmarks, and computational resources. We believe it is critical to establish a rigorous framework for evaluating these LLM-based zero-shot ranking approaches. By doing so, we can draw meaningful conclusions about their comparative effectiveness and efficiency.

Thus, in this paper, we first conduct a systematic evaluation of all existing approaches within a consistent experimental environment. In addition to assessing ranking effectiveness, we also compare the efficiency of these methods in terms of computational expense and query latency. Our findings indicate that the \textit{Pairwise} approach emerges as the most effective but falls short in terms of efficiency even with the assistance of sorting algorithms aimed at improving efficiency. Conversely, the \textit{Pointwise} approach stands out as the most efficient but lags behind other methods in terms of ranking effectiveness. 
The \textit{Listwise} approach, which relies solely on the generation of document labels in order, can strike a middle ground between efficiency and effectiveness but this varies considerably based on configuration, implementation and evaluation dataset (highlighting the importance of thoroughly evaluating these model under multiple settings).
Overall, these comprehensive results offer an understanding of the strengths and weaknesses of LLM-based zero-shot ranking approaches, providing valuable insights for those seeking to select the most suitable approach for real-world applications.

Having considered all the different approaches and their results in terms of efficiency and effectiveness tradeoffs, we set about devising a method that was both effective and efficient. Our approach was to take the most effective model (\textit{Pairwise}) and to enhance its efficiency (without seriously compromising effectiveness). Our solution is a novel \textit{Setwise} prompting approach. This concept stems from our realisation that the sorting algorithms employed by \textit{Pairwise} approaches can be accelerated by comparing multiple documents, as opposed to just a pair at a time.

Our \textit{Setwise} prompting approach instructs LLMs to select the most relevant document to the query from a set of candidate documents. This straightforward adjustment allows the sorting algorithms to infer relevance preferences for more than two candidate documents at each step, thus significantly reducing the total number of comparisons required; this leads to substantial savings in computational resources. Furthermore, beyond the adjustment to \textit{Pairwise} approaches, \textit{Setwise} prompting allows for the utilization of model output logits to estimate the likelihood of ranks of document labels, a capability not feasible in existing \textit{Listwise} approaches, which solely rely on document label ranking generation —-- a process that is slow and less effective. We \textit{Setwise} and other existing approaches under the same experimental settings to provide a clear and consistent comparison. Our results show that the incorporation of our \textit{Setwise} prompting substantially improves the efficiency of both \textit{Pairwise} and \textit{Listwise} approaches.
In addition, Setwise sorting enhances \textit{Pairwise} and \textit{Listwise} robustness to variations in the internal ordering quality of the initial rankings: no matter what the initial ordering of the top-k documents to rank is, our method provides consistent and effective results. This is unlike other methods that are highly susceptible to such initial ordering.

To conclude, this paper makes three key contributions to our understanding of LLM-based zero-shot ranking approaches:
\begin{enumerate}[leftmargin=14pt,itemsep=0mm]

	\item We introduce an innovative \textit{Setwise} prompting approach that enhances the sorting algorithms employed in the \textit{Pairwise} method, resulting in highly efficient zero-shot ranking with LLMs. 
	
	\item We conduct a systematic examination of all existing LLM-based zero-shot ranking approaches and our novel \textit{Setwise} approach under strict and consistent experimental conditions, including efficiency comparisons which have been overlooked in the literature. Our comprehensive empirical evaluation on popular zero-shot document ranking benchmarks offers valuable insights for practitioners.
	
	\item We further adapt how our \textit{Setwise} prompting approach computes rankings to the \textit{Listwise} approach, leveraging the model output logits to estimate the likelihood of rankings. This leads to a more effective and efficient \textit{Listwise} zero-shot ranking.

\end{enumerate}

\section{Background \& Related Work}

There are three main prompting approaches for zero-shot document ranking employing LLMs: \textit{Pointwise}~\cite{liang2022holistic,sachan-etal-2022-improving}, \textit{Listwise}~\cite{sun2023chatgpt,ma2023zero,pradeep2023rankvicuna}, and \textit{Pairwise}~\cite{qin2023large}. In this section, we delve into the specifics of these while situating our work within the existing literature. As a visual aid we will refer to Figure~\ref{fig:methods} as we discuss each method.

\vspace{-6pt}
\subsection{\textit{Pointwise} prompting approaches}
Figure~\ref{fig:methods}a shows pointwise approaches. 
There are two popular directions of prompting LLMs for ranking documents in a pointwise manner: \textit{generation} and \textit{likelihood}. In the generation approach, a ``yes/no" generation technique is used: LLMs are prompted to generate whether the provided candidate document is relevant to the query, with the process repeated for each candidate document. Subsequently, these candidate documents are re-ranked based on the normalized likelihood of generating a "yes" response~\cite{liang2022holistic,nogueira2020document}.
The likelihood approach involves query likelihood modelling (QLM)~\cite{ponte2017language,zhuang2021deep,zhuang2021tilde}, wherein LLMs are prompted to produce a relevant query for each candidate document. The documents are then re-ranked based on the likelihood of generating the actual query~\cite{sachan-etal-2022-improving}. 
It is worth noting that both pointwise methods require access to the output logits of the model to be able to compute the likelihood scores. Thus, it is not possible to use closed-sourced LLMs to implement these approaches if the corresponding APIs do not expose the logits values: this is the case for example of GPT-4. 

\vspace{-6pt}
\subsection{\textit{Listwise} prompting approaches}
Figure~\ref{fig:methods}b shows listwise approaches. 
Here the LLMs receive a query along with a list of candidate documents and are prompted to generate a ranked list of document labels based on their relevance to the query~\cite{sun2023chatgpt,ma2023zero,pradeep2023rankvicuna}. 
However, due to the limited input length allowed by LLMs, including all candidate documents in the prompt is not feasible. To address this, current listwise approaches use a sliding window method. This involves re-ranking a window of candidate documents, starting from the bottom of the original ranking list and progressing upwards. This process can be repeated multiple times to achieve an improved final ranking and allows for early stopping mechanisms to target only the top-$k$ ranking, thereby conserving computational resources. In contrast to pointwise methods, which utilize the likelihood value of the output tokens for ranking documents, listwise approaches rely on the more efficient process of generation of the ranking list.

\begin{figure*}[t]
	
	\begin{subfigure}{1\textwidth}
		\centering
		\includegraphics[width=\linewidth]{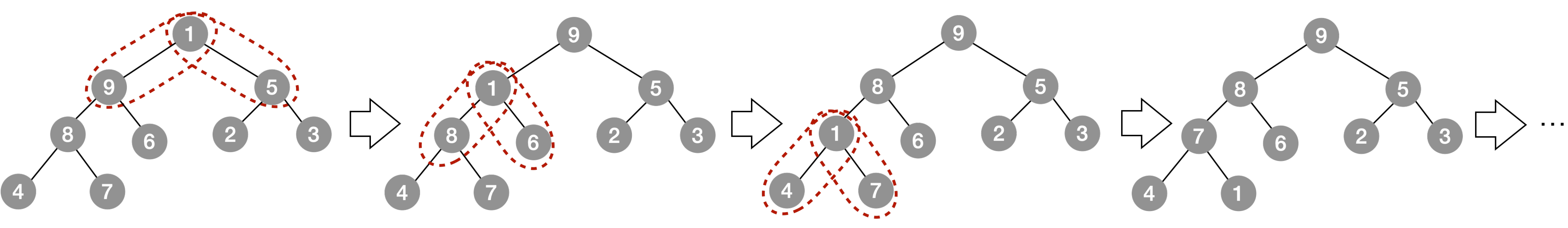}
		\caption{Heapify with Pairwise prompting  (comparing 2 documents at a time).}
		\label{fig:sfig1}
	\end{subfigure}
	\\
	\begin{subfigure}{1\textwidth}
		\centering
		\includegraphics[width=\linewidth]{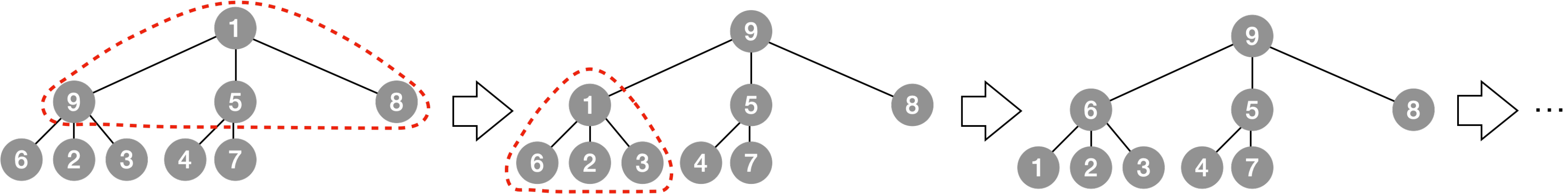}
		\caption{Heapify with our Setwise prompting (comparing 4 documents at a time).}
		\vspace{10pt}
		\label{fig:sfig2}
	\end{subfigure}
	\begin{subfigure}{1\textwidth}
		\includegraphics[width=\linewidth]{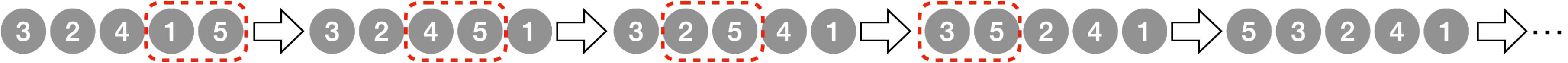}
		\caption{Bubble sort with Pairwise prompting  (comparing 2 documents at a time).}
			\vspace{10pt}
		\label{fig:sfig3}
	\end{subfigure}

	\begin{subfigure}{.6\textwidth}	
		\includegraphics[width=\linewidth]{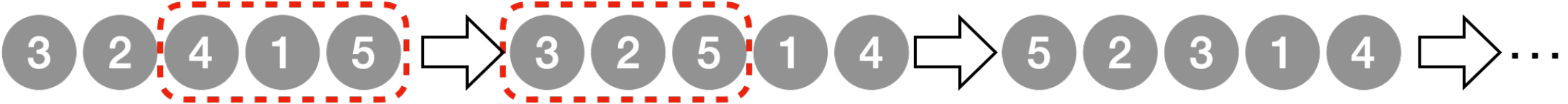}
		\caption{Bubble sort with our Setwise prompting (comparing 3 documents at a time).}
		\label{fig:sfig4}
	\end{subfigure}
\vspace{-6pt}
	\caption{Illustration of the impact of Setwise Prompting vs. Pairwise Prompting on Sorting Algorithms. Nodes are documents, numbers in nodes represent the level of relevance assigned by the LLM (higher is more relevant). \vspace{-6pt}}
	\label{fig:1}
\end{figure*}

\subsection{\textit{Pairwise} prompting approaches}
Figure~\ref{fig:methods}c shows pairwise approaches. LLMs are prompted with a query alongside a pair of documents, and are asked to generate the label indicating which document is more relevant to the  query~\cite{qin2023large,pradeep2021expando}. To re-rank all candidate documents, a basic method, called \textit{AllPairs}, involves generating all possible permutations of document pairs from the candidate set. Pairs are independently then fed into the LLM, and the preferred document for each pair is determined. Subsequently, an aggregation function is employed to assign a score to each document based on the inferred pairwise preferences, and the final ranking is established based on the total score assigned to each document~\cite{pradeep2021expando}. However, this aggregation-based approach suffers from high query latency: LLM inference on all document pairs can be computationally expensive. To address this efficiency issue in pairwise approaches, prior studies have introduced sampling~\cite{mikhailiuk2020active,10.1145/3539813.3545140} and sorting~\cite{qin2023large} algorithms. 
In this paper, we focus on sorting algorithms because, assuming an LLM can provide ideal pairwise preferences, the sorting algorithms offer the theoretical assurance of identifying the top-$k$ most relevant documents from the candidate pool. 
In prior work~\cite{qin2023large}, two sorting algorithms~\cite{knuth1997art}, \textit{heap sort} and \textit{bubble sort}, were employed. Unlike \textit{AllPairs}, these algorithms leverage efficient data structures to selectively compare document pairs, which can quickly pull the most relevant documents out from the candidate pool and place them at the top of the final ranking. This is particularly suitable for the top-$k$ ranking task, where only a ranking of the $k$ most relevant documents is needed. These sorting algorithms provide a stopping mechanism that prevents the need to rank all candidate documents.

\par From a theoretical standpoint the differences and relative advantages among these three families of zero-shot document ranking that employ LLMs are clear. However, from an empirical standpoint there has been no fair and comprehensive evaluation of these techniques in terms of effectiveness vs. efficiency, and across factors such as sizes of LLMs, benchmarks, and computational resources. 

\subsection{Other Directions in using LLMs for Ranking}
The three families of approaches outlined above directly use prompting of LLMs to infer document relevance to a given query, leveraging LLMs for document re-ranking in a zero-shot setting with no training data required for the ranking task. An alternative direction in using LLMs for retrieval has also emerged, where LLMs are instead used as text embedding models for dense document retrieval~\cite{lee2024gecko,wang2024text,wang2024improving,behnamghader2024llm2vec}. Because these methods create document representations independently of query representations, they can be used as a first-stage document retriever (i.e. in all effects as a bi-encoder architecture), rather than being limited to be used as a re-ranker as instead are the approaches from the three families we have reviewed above. However, these methods require performing contrastive learning training to enable the generative LLM to act as a text-embedding model. An exception to this is the recent PromptReps method~\cite{zhuang2024promptreps}, which does not require contrastive training, achieving document (and query) embedding simply via prompt engineering. Another feature of PromptReps is the ability to obtain at the same time both a dense and a sparse representation of documents and queries.

\section{Setwise Ranking Prompting}\label{sec:setwise}

In this section, we discuss the limitations present in the current LLM-based zero-shot ranking methods. We then describe our proposed \textit{Setwise} approach and how it addresses these limitations.
\subsection{Limitations of Current Approaches}

The efficiency of LLM-based zero-shot ranking methods hinges on two critical dimensions.

First, the number of LLM inferences significantly impacts efficiency. Given that LLMs are large neural networks with billions of parameters, inference is computationally intensive. Hence, an increased number of LLM inferences introduces a considerable computational overhead. This is notably observed in the current \textit{Pairwise} approach, which is inefficient due to the extensive need for inferring preferences for the many document pairs. While sorting algorithms offer some relief, they do not entirely mitigate the efficiency issue.

Second, the number of LLM-generated tokens per inference plays a pivotal role. LLMs employ a transformer decoder for autoregressive token generation, where the next token generation depend on previously tokens generated. Each additional generated token requires an extra LLM inference. This accounts for the inefficiency of the existing \textit{Listwise} approach, which relies on generating an entire ranking of document label lists, often requiring a substantial number of generated tokens.

Next, we introduce our new \textit{Setwise} prompting approaches, designed to overcome the efficiency limitations of both \textit{Pairwise} and \textit{Listwise} methods by minimizing the number of required LLM inferences and leveraging the logits produced by the LLM.

\begin{table}[t!]
	\centering
	\caption{
		Properties of different methods. Logits: requires access to the LLM logits. Generate: requires actually generating tokens. Batching: allows for batch inference. Top-$k$: allows for early stopping once top-$k$ most relevant documents found. \# LLM calls: the number of LLM forward passes needed in the worst case. ($N$: number of documents to re-rank. $r$: number of repeats.  $s$: step size for sliding window. $k$: number of top-$k$  documents to find. $c$: number of compared documents at each step.)
	\vspace{-6pt}}
	\resizebox{1\columnwidth}{!}{
		\begin{tabular}{lccccl}
			\toprule
			\bf Methods                                                      & \bf Logits                & \bf Generate            & \bf Batching  &  \bf Top-$k$    & \bf  \# LLM calls     \\
			\midrule
			\textit{pointwise.qlm}                                                & \cmark &                       &    \cmark &      &  $\mathcal{O}(N)$         \\
			\textit{pointwise.yes\_no}                                         & \cmark &                       &   \cmark   &      &  $\mathcal{O}(N)$         \\
			\textit{listwise.generation}                                          &                       & \cmark &   &  \cmark  &  $\mathcal{O}(r * (N/s))$              \\
			\textit{listwise.likelihood}                                          & \cmark &                       &      &  \cmark  & $\mathcal{O}(r * (N/s))$            \\
			\textit{pairwise.allpair}                                             & \cmark & \cmark & \cmark &     &$\mathcal{O}(N^2 - N) $ \\
			\textit{pairwise.heapsort}                                            & \cmark & \cmark &        &  \cmark    &$\mathcal{O}(k * \log_{2}{N})$         \\
			\textit{pairwise.bubblesort}                                          & \cmark & \cmark &       &  \cmark     &$\mathcal{O}(k * N )$         \\
			\textit{setwise.heapsort}                                             & \cmark & \cmark &          & \cmark  &$\mathcal{O}(k *  \log_{c}{N})$         \\
			\textit{setwise.bubblesort}                                           & \cmark & \cmark &         & \cmark    &$\mathcal{O}(k * (N/(c-1)))$         \\
			\bottomrule
			\vspace{-20pt}
		\end{tabular}
		\label{tab:features}
	}
\end{table}

\vspace{-6pt}
\subsection{Speeding-up Pairwise with Setwise}
To solve the inefficiency issue of these approaches, we propose a novel \textit{Setwise} prompting approach. Our prompt, as illustrated in Figure~\ref{fig:methods}d, instructs the LLM to select the most relevant document for the given query from a set of documents, hence the term \textit{Setwise} prompting. We specifically treat the collection of documents as an unordered set and later experiments will show that \textit{Setwise} prompting is quite robust to document ordering.

With our prompt, sorting-based \textit{Pairwise} approaches can be considerably accelerated. This is because the original \textit{heap sort} and \textit{bubble sort} algorithm used in the \textit{Pairwise} approach only compares a pair of documents at each step in the sorting process, as illustrated in Figure~\ref{fig:sfig1} and \ref{fig:sfig3}. These sorting algorithms can be sped up by comparing more than two documents at each step. 
For example, in the \textit{heap sort} algorithm, the ``heapify" function needs to be invoked for each subtree, where the parent node must be swapped with the child node with the highest value if it exceeds the parent value. In the case of Figure~\ref{fig:sfig1}, to perform ``heapify" with pairwise prompting, a minimum of 6 comparisons (each root node paired with each child node) are required. Conversely, if we increase the number of child nodes in each subtree to 3 and can compare 4 nodes at a time, only 2 comparisons are needed to ``heapify" a tree with 9 nodes, as illustrated in Figure~\ref{fig:sfig2}. Similarly, for the \textit{bubble sort} algorithm, if we can compare more than a pair of documents at a time, each ``bubbling'' process will be accelerated. For instance, in Figure~\ref{fig:sfig3}, there are 4 comparisons in total, but in Figure~\ref{fig:sfig4}, with the ability to compare 3 documents at once, only 2 comparisons are required to be able to bring the node with the largest value to the top. Our \textit{Setwise} prompting is designed to instruct LLMs to compare the relevance of multiple documents at a time, making it well-suited for this purpose.

\vspace{-6pt}
\subsection{Listwise Likelihoods with Setwise}

Our \textit{Setwise} prompting can also accelerate the ranking process for the \textit{Listwise} approach. The original \textit{Listwise} method relies on the LLM's next token generation to produce the complete ordered list of document labels at each step of the sliding window process, as illustrated in Figure~\ref{fig:methods}b. As we discussed, generating the document label list is computationally intensive, because the LLM must do one inference for each next token prediction. On the other hand, the LLM may generate results in an unexpected format or even decline to generate the desired document label list~\cite{sun2023chatgpt}, thus harming effectiveness. Fortunately, if we have access to the LLM's output logits, these issues can be avoided by evaluating the likelihood of generating every conceivable document label list and then selecting the most probable one as the output. Regrettably, this is only theoretically possible, but in practice, it is unfeasible for the existing \textit{Listwise} approach due to the very large number of possible document label permutation, which implies that the process of likelihood checking may actually become even more time-consuming than generating the list itself.

\textit{Setwise} prompting again provides a solution: we can easily derive an ordered list of document labels from the LLM output logits. This is done by assessing the likelihood of each document label being chosen as the most relevant, as shown in Figure~\ref{fig:methods}d. This straightforward trick markedly accelerates \textit{Listwise} ranking, as it requires only a single forward pass of the LLM, and also guarantees that the output matches the desired document label list.

\vspace{-6pt}
\subsection{Advantages of Setwise}

We summarize and compare the key different properties of existing zero-shot LLM ranking approaches along with our proposed \textit{Setwise} prompting approach in Table~\ref{tab:features}. Notably, \textit{pointwise.qlm}, \textit{pointwise.yes\_no} and \textit{pairwise.allpair} require a brute-force of LLM inference for all available documents relevance or preferences. Thus, they are unable to facilitate early-stopping for the top-$k$ ranking. However, these approaches do allow batch inferences, hence the maximum GPU memory utilization could be easily achieved by using the highest batch size. On the other hand, other approaches use sorting algorithms, enabling early-stopping once the top-$k$ most relevant documents are identified. However, this compromises the feasibility of batching inference, as the LLM inference at each step of the sorting algorithms relies on the results from the preceding step. Our \textit{Setwise} prompting empowers the previous \textit{Listwise} approach (\textit{listwise.generation}), which relied on LLM's next token generations, to now utilize the LLM's output logits. We refer to the \textit{Listwise} approach that incorporates our \textit{Setwise} prompt as \textit{listwise.likelihood}. Finally, comparing with \textit{Pairwise} approaches, our \textit{Setwise} prompting has fewer LLM calls by comparing a minimum of $c \geq 3$ documents at each step of the sorting algorithms.

On the other hand, model output calibration might be a concern for \textit{Pointwise} methods because each document's relevance is inferred independently. Consequently, ranking documents based on \textit{Pointwise} relevance scores necessitates calibration~\cite{qin2023large}. However, for our \textit{Setwise} method (as well as for \textit{Pairwise} and \textit{Listwise}) the model output logits are derived from the input of multiple documents, which do not directly represent the relevance of a single document but rather serve as an indicator of preference among documents. Thus, calibration is not necessary.

\section{Experiments}

\subsection{Datasets and evaluations}
The first objective of this study is to contribute a fair and comprehensive evaluation of existing LLM-based zero-shot ranking methods in terms ranking effectiveness and efficiency. To achieve this goal, we carried out extensive empirical evaluations using well-established document ranking datasets: the TREC Deep Learning   2019~\cite{craswell2020overview} and 2020~\cite{craswell2021overview}, along with the BEIR benchmark datasets~\cite{thakur2021beir}. 
To guarantee a fair comparison across different approaches, we tested all of the methods using the same open-source Flan-t5 LLMs~\cite{wei2021finetuned}, available on the Huggingface model hub in various sizes (780M, 3B, and 11B parameters). All LLM methods were used to re-rank 100 documents retrieved by a BM25 first-stage retriever. 
In order to optimize efficiency, the focus was on a top-$k$ ranking task, whereby the re-ranking process stopped as soon as the top-$k$ most relevant documents were identified and ranked. Here, we set $k=10$. The effectiveness of different approaches was evaluated using the NDCG@10 metric, which serves as the official evaluation metric for the employed datasets.

Efficiency was evaluated with the following metrics:
\begin{itemize}
	\item \textit{The average number of LLM inferences per query.} LLMs have limited input length. Thus, to re-rank 100 documents, multiple LLM inferences are often needed. It's important to note that an increased number of LLM inferences translates to higher computational demands. Thus, we regard this as an efficiency metric worth considering.
	\item \textit{The average number of prompt tokens inputted to the LLMs per query.} This metric takes into account the actual average quantity of input tokens required in the prompts for each method to re-rank 100 documents per query. Given that self-attention mechanisms in transformer-based LLMs become prohibitively costly for a large number of input tokens~\cite{10.5555/3295222.3295349}, an increase in tokens within the prompts also translates to higher computational demands. Notably, numerous LLM web API services, including OpenAI APIs, charge based on the number of input tokens in the API calls. As such, we deem this metric valuable in assessing efficiency.
	\item \textit{The average number of generated tokens outputted by LLMs per query.} Much like the assessment of average prompt tokens, this metric provides an evaluation of computational efficiency, but from a token generation perspective. Instead of focusing on the number of tokens in the prompt, it takes into account the number of tokens generated. This is particularly significant because transformer-based generative LLMs produce content token-by-token, with each subsequent token relying on the generation of preceding ones. Consequently, an increase in number of generated tokens leads to a corresponding increase in the computational cost, as each additional generated token implies another LLM forward inference. In fact, OpenAI applies a pricing structure wherein the cost for the number of generated tokens is twice that of the number of prompt tokens for their LLM APIs~\footnote{\url{https://openai.com/pricing}, last visited 12 October 2023.}. This underscores the substantial impact that generated tokens can have on computational expenses.
	\item \textit{The average query latency.} We evaluate the run time efficiency of all the methods with average query latency. To conduct this assessment, a single GPU is employed, and queries are issued one at a time. The per-query latency is then averaged across all the queries in the dataset. It's important to highlight that for methods that support batching we always employ the maximum batch size to optimize GPU memory usage and parallel computation, thus maximizing efficiency for these particular methods. This approach ensures that the evaluation is conducted under conditions most favourable for efficiency gains. It is important to acknowledge that while other methods may not be able to use the batching strategy for individual queries, they do have the capability to utilize batching and parallel computing across various user queries in real-world scenarios. However, this lies more in engineering efforts and falls outside the scope of this paper: as such, we do not investigate this perspective.
\end{itemize}

\begin{table*}[]
	\centering
	\caption{
		Results on TREC DL. All the methods re-rank BM25 top 100 documents. We present the ranking effectiveness in terms of NDCG@10, best values highlighted in boldface.
		Superscripts denote statistically significant improvements (paired Student's t-test with $p \le 0.05$ with Bonferroni correction).
		\#Inferences denotes the average number of LLM inferences per query. Pro. Tokens is the average number of tokens in the prompt for each query. Gen. tokens is the average number of generated tokens per query. Latency is the average query latency, in seconds.
	}

	\resizebox{1\textwidth}{!}{
		\begin{tabular}{cc|l|lllll|lllll}
			\toprule
			\multicolumn{3}{c}{} & \multicolumn{5}{|c|}{\textbf{TREC DL 2019}} &  \multicolumn{5}{|c}{\textbf{TREC DL 2020}}
			\\
			\toprule
			& {\small \textbf{\#}}
			&{\small \textbf{Methods}}
			& {\small \textbf{NDCG@10}}
			& {\small \textbf{\#Inferences}}
			&  {\small \textbf{Pro. tokens}}
			& {\small \textbf{Gen. tokens}}
			& {\small \textbf{Latency(s)}}
			& {\small \textbf{NDCG@10}}
			& {\small \textbf{\#Inferences}}
			&  {\small \textbf{Pro. tokens}}
			& {\small \textbf{Gen. tokens}}
			& {\small \textbf{Latency(s)}} \\ 
			\midrule
			
			& $a$ &
			BM25 &
			.506\hphantom{$^{bcdefghij}$} & - & -& -& -& .480\hphantom{$^{bcdefghij}$}& -& -& -& -\\
			\bottomrule
			\multirow{9}{*}{ \rotatebox[origin=c]{90}{Flan-t5-large}}
			&$b$ &
			pointwise.qlm &
			.557\hphantom{$^{acdefghij}$} & 100 & 15211.6 & - & 0.6  & .567$^{a}$\hphantom{$^{cdefghij}$} & 100 & 15285.2 & - & 0.5  \\
			&$c$ &
			pointwise.yes\_no &
			.654$^{abd}$\hphantom{$^{efghij}$} & 100 & 16111.6 & - & 0.6 & .615$^{ad}$\hphantom{$^{befghij}$} & 100 & 16185.2 & - & 0.6 \\
			&$d$ &
			listwise.generation &
			.561$^{a}$\hphantom{$^{bcefghij}$}& 245 & 119120.8 & 2581.35 & 54.2& .547$^{a}$\hphantom{$^{bcefghij}$} & 245 & 119629.6 & 2460.1 & 52  \\
			&$e$ &
			listwise.likelihood &
			.669$^{abd}$\hphantom{$^{cfghij}$} & 245 & 94200.7 & - & 10 &\textbf{.626}$^{abd}$\hphantom{$^{cfghij}$} & 245 & 95208.3 & - &  10 \\
			& $f$ &
			pairwise.allpair & 
			.666$^{abd}$\hphantom{$^{ceghij}$} & 9900 & 3014383.1 & 49500 & 109.6 &  .622$^{abd}$\hphantom{$^{ceghij}$} & 9900 & 3014232.7 & 49500 &  108.9  \\
			&$g$ &
			pairwise.heapsort &
			.657$^{abd}$\hphantom{$^{cefhij}$} & 230.3 & 104952.5 & 2303.3 & 16.1 &  .619$^{abd}$\hphantom{$^{cefhij}$} & 226.8 & 104242.1 &2268.3 & 16.1  \\
			&$h$ &
			pairwise.bubblesort &
			.636$^{abd}$\hphantom{$^{cefgij}$} & 844.2 & 381386.3 & 8441.6 & 58.3 & .589$^{ad}$\hphantom{$^{bcefgij}$} & 778.5 & 357358.5&7785.4 &54.1 \\
			&$i$ &
			setwise.heapsort &
			.670$^{abd}$\hphantom{$^{cefghj}$} & 125.4 & 40460.6 & 626.9 & 8.0 & .618$^{ad}$\hphantom{$^{bcefghj}$} & 124.2 &40362.0 & 621.1& 8.0 \\
			&$j$ &
			setwise.bubblesort &
			\textbf{.678}$^{abdh}$\hphantom{$^{cefgj}$} & 460.5 & 147774.1 & 2302.3 & 29.1 & .624$^{abd}$\hphantom{$^{cehfgi}$} &457.4 &148947.3& 2287.1&28.9\\
			\bottomrule
			\multirow{9}{*}{ \rotatebox[origin=c]{90}{Flan-t5-xl}}
			& $b$ &
			pointwise.qlm &
			.542\hphantom{$^{acdefghij}$} & 100 & 15211.6 & - & 1.4 & .542$^{a}$\hphantom{$^{cdefghij}$} & 100&15285.2 & -& 1.4  \\
			&$c$ &
			pointwise.yes\_no &
			.650$^{abd}$\hphantom{$^{efghij}$} & 100 & 16111.6& - & 1.5 &  .636$^{abd}$\hphantom{$^{efghij}$} &100 &16185.2 & - &1.5 \\
			&$d$ &
			listwise.generation &
			.569$^{a}$\hphantom{$^{bcefghij}$} & 245 &  119163.0 & 2910 & 71.4 &  .547$^{a}$\hphantom{$^{bcefghij}$} &245 &119814.3 &2814.7 & 69 \\
			&$e$ &
			listwise.likelihood &
			.689$^{abd}$\hphantom{$^{cfghij}$} & 245 & 94446.1 & - & 12.5 &  .672$^{abd}$\hphantom{$^{cfghij}$} &245 & 95298.7& - &12.6  \\
			&$f$ &
			pairwise.allpair &
			\textbf{.713}$^{abcdehi}$\hphantom{$^{gj}$}  & 9900 & 2953436.2 & 49500 & 254.9 &  .682$^{abcd}$\hphantom{$^{eghij}$ } & 9900& 2949457.6& 49500&254.8 \\
			&$g$&
			pairwise.heapsort &
			.705$^{abcd}$\hphantom{$^{efhij}$} & 241.9 & 110126.9 & 2418.6 & 20.5 &  \textbf{.692}$^{abcdh}$\hphantom{$^{efij}$} &244.3 &111341 &2443.3 &20.8 \\
			&$h$&
			pairwise.bubblesort &
			.683$^{abd}$\hphantom{$^{cefgij}$} & 886.9 & 400367.1 & 8869.1 & 75.1 &  .662$^{abd}$\hphantom{$^{cefgij}$} & 863.9 & 394954.2 & 8638.5 &74.3\\
			&$i$&
			setwise.heapsort &
			.693$^{abcd}$\hphantom{$^{efghj}$} & 129.5 & 41665.7 & 647.4 & 9.6 &  .678$^{abcd}$\hphantom{$^{efghj}$} & 127.8 & 41569.1& 638.9&9.7 \\
			&$j$&
			setwise.bubblesort &
			.705$^{abcd}$\hphantom{$^{efghi}$} & 466.9 & 149949.1 & 2334.5 & 35.2 &  .676$^{abcd}$\hphantom{$^{efghi}$} & 463.5 & 151249.8& 2317.6& 35.3 \\
			\bottomrule
			\multirow{9}{*}{ \rotatebox[origin=c]{90}{Flan-t5-xxl}} 
			&$b$ &
			pointwise.qlm &
			.506\hphantom{$^{acdefghij}$} & 100 & 15211.6 & - & 3.7 &  .492\hphantom{$^{acdefghij}$} &100& 15285.2& -& 3.7\\
			&$c $&
			pointwise.yes\_no &
			.644$^{ab}$\hphantom{$^{defghij}$} & 100 & 16111.6& - & 3.9 &  .632$^{ab}$\hphantom{$^{defghij}$} & 100&16185.2 &- & 3.9 \\
			&$d$ &
			listwise.generation &
			.662$^{ab}$\hphantom{$^{cefghij}$} & 245 & 119334.7 & 2824 & 100.1 &  .637$^{ab}$\hphantom{$^{cefghij}$} & 245& 119951.6& 2707.9& 97.3\\
			&$e$ &
			listwise.likelihood &
			.701$^{abcd}$\hphantom{$^{fghji}$} & 245 & 94537.5 & - & 36.6 &  .690$^{abcd}$\hphantom{$^{fghij}$} & 245& 95482.7& -& 36.9\\
			& $f $&
			pairwise.allpair &
			.699$^{abcd}$\hphantom{$^{eghij}$} & 9900 & 2794942.6& 49500 & 730.2 &  .688$^{abcd}$\hphantom{$^{eghij}$}&9900 & 2794928.4& 49500& 730.5\\
			&$g$ &
			pairwise.heapsort &
			.708$^{abcdh}$\hphantom{$^{efij}$} & 239.4 & 109402 & 2394 & 45 &  \textbf{.699}$^{abcd}$\hphantom{$^{efhij}$} & 240.5& 110211.8& 2404.8& 45.2\\
			&h &
			pairwise.bubblesort &
			.679$^{ab}$\hphantom{$^{cdefgij}$} & 870.5 & 394386 & 8705.3 & 162.5 &  .681$^{abcd}$\hphantom{$^{efgij}$} & 842.9& 387359.2& 8428.5& 158.8\\
			&i &
			setwise.heapsort &
			.706$^{abcd}$\hphantom{$^{efghj}$} & 130.1 & 42078.6 & 650.5 & 20.2 &  
			.688$^{abcd}$\hphantom{$^{efghj}$} & 128.1& 41633.7& 640.6& 20.0\\
			&$j$ &
			setwise.bubblesort &
			\textbf{.711}$^{abcdh}$\hphantom{$^{efgi}$} & 468.3& 150764.8& 2341.6 & 72.6 &  .686$^{abcd}$\hphantom{$^{efghi}$} & 467.9& 152709.5& 2339.6&73.2\\
			\bottomrule
		\end{tabular}
	}
	\label{tab:results}
\end{table*}

\subsection{Implementation details}

To establish the initial BM25 first-stage ranking for all datasets, we employed the Pyserini Python library~\cite{10.1145/3404835.3463238} with default settings. For LLM-based zero-shot re-rankers, we followed the prompts recommended in existing literature to guide Flan-t5 models of varying sizes (Flan-t5-large with 780M parameters, Flan-t5-xl with 3B parameters, and Flan-t5-xxl with 11B parameters) in executing the zero-shot ranking task.

Specifically, for the \textit{pointwise.qlm} method, we adopted the prompt suggested by~\citet{sachan-etal-2022-improving}. For \textit{pointwise.yes\_no}, we use the prompt provided by~\citet{qin2023large}. For \textit{listwise.generate}, we utilized the prompt designed by~\citet{sun2023chatgpt}. As for \textit{pairwise.allpair}, \textit{pairwise.heapsort}, and \textit{pairwise.bubblesort}, we relied on the prompts from the original paper by~\citet{qin2023large}. 
For methods leveraging our \textit{Setwise} prompting (i.e. \textit{listwise.likelihood}, \textit{setwise.heapsort}, and \textit{setwise.bubblesort}), we employed the prompts detailed in Section~\ref{sec:setwise}.

In the case of \textit{Listwise} approaches, we configure the window size ($w$) to contain 4 documents, each capped at a maximum of 100 tokens. The step size ($s$) is set to 2, and the number of repetitions ($r$) is set to 5. 
These settings take into account the token limitations imposed by Flan-t5 models, which have an input token cap of 512. A window size of 4 documents appears reasonable as it aligns well with the prompt capacity. Additionally, a step size of 2, combined with 5 repetitions, has theoretical guarantees of bringing the 10 most relevant documents to the top. 
For our \textit{Setwise} approaches, we set the number of compared documents $c$ in each step to 3 for the main results. We further investigate the impact of $c$ in Section~\ref{sec:trade-off}. For all other methods, we truncate the documents with the maximum number of tokens to 128. 

We note that, among all the methods capable of utilizing both model output logits and generation outputs, we exclusively employ the latter. This choice is made in favor of a more general approach that allows for leveraging generation APIs across a wider range of closed-source LLMs. Nevertheless, we investigate the difference between using model output logits and generation outputs for our \textit{Setwise} approaches in Section~\ref{sec:effectiveness}.
 
We carried out the efficiency evaluations on a local GPU workstation equipped with an AMD Ryzen Threadripper PRO 3955WX 16-Core CPU, a NVIDIA RTX A6000 GPU with 49GB of memory, and 128GB of DDR4 RAM.

\begin{table*}[]
	\centering
	\caption{
		Overall NDCG@10 obtained by  methods on BEIR datasets.
		The best results are highlighted in boldface.
		Superscripts denote statistically significant improvements (paired Student's t-test with $p \le 0.05$ with Bonferroni correction).
	}
	\resizebox{1\textwidth}{!}{
		\begin{tabular}{cc|l|llllllll|l}
			\toprule
			&\textbf{\#}
			& \textbf{Methods}
			& \textbf{Covid}
			&\textbf{NFCorpus}
			&\textbf{Touche}
			&\textbf{DBPedia}
			&\textbf{SciFact}
			&\textbf{Signal}
			&\textbf{News}
			&\textbf{Robust04}
			& \textbf{Avg}
			\\ 
			\midrule
			& $a$ & BM25 & .595 & .322 & .442 & .318 & .679& .331 & .395 & .407 & .436\\
			
			\toprule
			\multirow{8}{*}{ \rotatebox[origin=c]{90}{Flan-t5-large}}
			
			& $b$ & pointwise.qlm & .664$^{a}$\hphantom{$^{cdefghi}$} & .322\hphantom{$^{acdefghi}$} & .260\hphantom{$^{acdefghi}$} & .305\hphantom{$^{acdefghi}$} & .644$^{c}$\hphantom{$^{adefghi}$} & .314$^{c}$\hphantom{$^{adefghi}$} & .413$^{c}$\hphantom{$^{adefghi}$} & .439$^{af}$\hphantom{$^{cdeghi}$} & .420\\
			& $c$ & pointwise.yes\_no & .664$^{a}$\hphantom{$^{bdefghi}$} & .308\hphantom{$^{abdefghi}$} & .238\hphantom{$^{abdefghi}$} & .296\hphantom{$^{abdefghi}$} & .504\hphantom{$^{abdefghi}$} & .275\hphantom{$^{abdefghi}$} & .346\hphantom{$^{abdefghi}$} & .456$^{af}$\hphantom{$^{bdeghi}$} & .386\\
			& $d$ & listwise.generation & .692$^{a}$\hphantom{$^{bcefghi}$} & .333$^{ac}$\hphantom{$^{befghi}$} & .441$^{bcefhi}$\hphantom{$^{ag}$} & .391$^{abc}$\hphantom{$^{efghi}$} & .650$^{c}$\hphantom{$^{abefghi}$} & .343$^{ace}$\hphantom{$^{bfghi}$} & .428$^{ac}$\hphantom{$^{befghi}$} & .441$^{af}$\hphantom{$^{bceghi}$} & .465\\
			& $e$ & listwise.likelihood & .756$^{abcd}$\hphantom{$^{fghi}$} & .334$^{c}$\hphantom{$^{abdfghi}$} & .327$^{bc}$\hphantom{$^{adfghi}$} & \textbf{.444}$^{abcdfgh}$\hphantom{$^{i}$} & .639$^{c}$\hphantom{$^{abdfghi}$} & .308$^{c}$\hphantom{$^{abdfghi}$} & \textbf{.453}$^{ac}$\hphantom{$^{bdfghi}$} & .475$^{abdfg}$\hphantom{$^{chi}$} & .467\\
			& $f$ & pairwise.heapsort & .761$^{abcdg}$\hphantom{$^{ehi}$} & .336$^{c}$\hphantom{$^{abdeghi}$} & .318$^{bc}$\hphantom{$^{adeghi}$} & .414$^{abcd}$\hphantom{$^{eghi}$} & .671$^{chi}$\hphantom{$^{abdeg}$} & .325$^{c}$\hphantom{$^{abdeghi}$} & .440$^{ac}$\hphantom{$^{bdeghi}$} & .402\hphantom{$^{abcdeghi}$} & .458\\
			& $g$ & pairwise.bubblesort & .714$^{a}$\hphantom{$^{bcdefhi}$} & \textbf{.341}$^{abcdh}$\hphantom{$^{efi}$} & \textbf{.447}$^{bcefhi}$\hphantom{$^{ad}$} & .416$^{abcd}$\hphantom{$^{efhi}$} & \textbf{.700}$^{bcdefhi}$\hphantom{$^{a}$} & \textbf{.361}$^{abcdefh}$\hphantom{$^{i}$} & .440$^{ac}$\hphantom{$^{bdefhi}$} & .439$^{af}$\hphantom{$^{bcdehi}$} & .482\\
			& $h$ & setwise.heapsort & \textbf{.768}$^{abcdg}$\hphantom{$^{efi}$} & .325$^{c}$\hphantom{$^{abdefgi}$} & .303$^{c}$\hphantom{$^{abdefgi}$} & .413$^{abcd}$\hphantom{$^{efgi}$} & .620$^{c}$\hphantom{$^{abdefgi}$} & .319$^{c}$\hphantom{$^{abdefgi}$} & .439$^{c}$\hphantom{$^{abdefgi}$} & .462$^{abf}$\hphantom{$^{cdegi}$} & .456\\
			&$ i$ & setwise.bubblesort & .761$^{abcdg}$\hphantom{$^{efh}$} & .338$^{abch}$\hphantom{$^{defg}$} & .394$^{bcefh}$\hphantom{$^{adg}$} & .441$^{abcdfgh}$\hphantom{$^{e}$} & .636$^{c}$\hphantom{$^{abdefgh}$} & .351$^{bcefh}$\hphantom{$^{adg}$} & .447$^{ac}$\hphantom{$^{bdefgh}$} & \textbf{.497}$^{abcdefgh}$\hphantom{} & \textbf{.483}\\
			
			\bottomrule
			\multirow{8}{*}{ \rotatebox[origin=c]{90}{Flan-t5-xl}}
			& $b$ & pointwise.qlm & .679$^{a}$\hphantom{$^{cdefghi}$} & .330\hphantom{$^{acdefghi}$} & .216\hphantom{$^{acdefghi}$} & .310$^{c}$\hphantom{$^{adefghi}$} & .696$^{c}$\hphantom{$^{adefghi}$} & .299\hphantom{$^{acdefghi}$} & .422\hphantom{$^{acdefghi}$} & .427\hphantom{$^{acdefghi}$} & .422\\
			& $c$ & pointwise.yes\_no & .698$^{a}$\hphantom{$^{bdefghi}$} & .331\hphantom{$^{abdefghi}$} & .269\hphantom{$^{abdefghi}$} & .273\hphantom{$^{abdefghi}$} & .553\hphantom{$^{abdefghi}$} & .297\hphantom{$^{abdefghi}$} & .413\hphantom{$^{abdefghi}$} & .479$^{ab}$\hphantom{$^{defghi}$} & .414\\
			& $d$ & listwise.generation & .650$^{a}$\hphantom{$^{bcefghi}$} & .334$^{a}$\hphantom{$^{bcefghi}$} & \textbf{.451}$^{bcefghi}$\hphantom{$^{a}$} & .366$^{abc}$\hphantom{$^{efghi}$} & .694$^{c}$\hphantom{$^{abefghi}$} & .349$^{abcefh}$\hphantom{$^{gi}$} & .437$^{a}$\hphantom{$^{bcefghi}$} & .475$^{ab}$\hphantom{$^{cefghi}$} & .470\\
			& $e$ & listwise.likelihood & .736$^{abd}$\hphantom{$^{cfghi}$} & \textbf{.360}$^{abcd}$\hphantom{$^{fghi}$} & .310$^{b}$\hphantom{$^{acdfghi}$} & \textbf{.449}$^{abcdfghi}$\hphantom{} & .686$^{c}$\hphantom{$^{abdfghi}$} & .320\hphantom{$^{abcdfghi}$} & .472$^{abc}$\hphantom{$^{dfghi}$} & .526$^{abcd}$\hphantom{$^{fghi}$} & .482\\
			& $f$ & pairwise.heapsort & \textbf{.778}$^{abcde}$\hphantom{$^{ghi}$} & .355$^{abcd}$\hphantom{$^{eghi}$} & .303$^{b}$\hphantom{$^{acdeghi}$} & .417$^{abcd}$\hphantom{$^{eghi}$} & .711$^{ch}$\hphantom{$^{abdegi}$} & .317\hphantom{$^{abcdeghi}$} & .471$^{abc}$\hphantom{$^{deghi}$} & .550$^{abcdehi}$\hphantom{$^{g}$} & .488\\
			& $g$ & pairwise.bubblesort & .763$^{abcd}$\hphantom{$^{efhi}$} & .359$^{abcd}$\hphantom{$^{efhi}$} & .400$^{bcefhi}$\hphantom{$^{ad}$} & .432$^{abcdf}$\hphantom{$^{ehi}$} & \textbf{.734}$^{abcdefhi}$\hphantom{} & .353$^{bcefh}$\hphantom{$^{adi}$} & .485$^{abcd}$\hphantom{$^{efhi}$} & \textbf{.553}$^{abcdehi}$\hphantom{$^{f}$} & \textbf{.510}\\
			& $h$ & setwise.heapsort & .757$^{abcd}$\hphantom{$^{efgi}$} & .352$^{abcd}$\hphantom{$^{efgi}$} & .283$^{b}$\hphantom{$^{acdefgi}$} & .428$^{abcdf}$\hphantom{$^{egi}$} & .677$^{c}$\hphantom{$^{abdefgi}$} & .314\hphantom{$^{abcdefgi}$} & .465$^{ac}$\hphantom{$^{bdefgi}$} & .520$^{abcd}$\hphantom{$^{efgi}$} & .475\\
			& $i$ & setwise.bubblesort & .756$^{abcd}$\hphantom{$^{efgh}$} & .353$^{abcd}$\hphantom{$^{efgh}$} & .330$^{bch}$\hphantom{$^{adefg}$} & .438$^{abcdfh}$\hphantom{$^{eg}$} & .691$^{c}$\hphantom{$^{abdefgh}$} & \textbf{.362}$^{abcefh}$\hphantom{$^{dg}$} & \textbf{.497}$^{abcd}$\hphantom{$^{efgh}$} & .537$^{abcdh}$\hphantom{$^{efg}$} & .496\\
			
			\bottomrule
			\multirow{8}{*}{ \rotatebox[origin=c]{90}{Flan-t5-xxl}}
			& $b$ & pointwise.qlm & .707$^{a}$\hphantom{$^{cdefghi}$} & .342$^{ac}$\hphantom{$^{defghi}$} & .188\hphantom{$^{acdefghi}$} & .324\hphantom{$^{acdefghi}$} & .712$^{c}$\hphantom{$^{adefghi}$} & .307$^{c}$\hphantom{$^{adefghi}$} & .431\hphantom{$^{acdefghi}$} & .440$^{a}$\hphantom{$^{cdefghi}$} & .431\\
			& $c$ & pointwise.yes\_no & .691$^{a}$\hphantom{$^{bdefghi}$} & .322\hphantom{$^{abdefghi}$} & .240$^{b}$\hphantom{$^{adefghi}$} & .305\hphantom{$^{abdefghi}$} & .623\hphantom{$^{abdefghi}$} & .274\hphantom{$^{abdefghi}$} & .392\hphantom{$^{abdefghi}$} & .515$^{ab}$\hphantom{$^{defghi}$} & .420\\
			& $d$ & listwise.generation & .664$^{a}$\hphantom{$^{bcefghi}$} & .344$^{ac}$\hphantom{$^{befghi}$} & \textbf{.453}$^{bcefhi}$\hphantom{$^{ag}$} & \textbf{.441}$^{abcefghi}$\hphantom{} & .736$^{ac}$\hphantom{$^{befghi}$} & .353$^{bcefh}$\hphantom{$^{agi}$} & .458$^{ac}$\hphantom{$^{befghi}$} & .495$^{ab}$\hphantom{$^{cefghi}$} & .493\\
			& $e$ & listwise.likelihood & .749$^{acd}$\hphantom{$^{bfghi}$} & .352$^{ac}$\hphantom{$^{bdfghi}$} & .307$^{bc}$\hphantom{$^{adfghi}$} & .416$^{abch}$\hphantom{$^{dfgi}$} & .725$^{ac}$\hphantom{$^{bdfghi}$} & .316$^{c}$\hphantom{$^{abdfghi}$} & .479$^{abc}$\hphantom{$^{dfghi}$} & .518$^{abd}$\hphantom{$^{cfghi}$} & .483\\
			& $f$ & pairwise.heapsort & .738$^{acd}$\hphantom{$^{beghi}$} & .359$^{abcdhi}$\hphantom{$^{eg}$} & .324$^{bc}$\hphantom{$^{adeghi}$} & .407$^{abc}$\hphantom{$^{deghi}$} & .744$^{abc}$\hphantom{$^{deghi}$} & .328$^{c}$\hphantom{$^{abdeghi}$} & .487$^{abc}$\hphantom{$^{deghi}$} & .543$^{abcdeh}$\hphantom{$^{gi}$} & .491\\
			& $g$ & pairwise.bubblesort & .733$^{ad}$\hphantom{$^{bcefhi}$} & \textbf{.363}$^{abcdehi}$\hphantom{$^{f}$} & .423$^{bcefh}$\hphantom{$^{adi}$} & .421$^{abcfh}$\hphantom{$^{dei}$} & \textbf{.756}$^{abcdeh}$\hphantom{$^{fi}$} & \textbf{.355}$^{bcefh}$\hphantom{$^{adi}$} & \textbf{.490}$^{abcd}$\hphantom{$^{efhi}$} & \textbf{.550}$^{abcdehi}$\hphantom{$^{f}$} & \textbf{.511}\\
			& $h$ & setwise.heapsort & .752$^{abcd}$\hphantom{$^{efgi}$} & .346$^{ac}$\hphantom{$^{bdefgi}$} & .297$^{bc}$\hphantom{$^{adefgi}$} & .402$^{abc}$\hphantom{$^{defgi}$} & .726$^{ac}$\hphantom{$^{bdefgi}$} & .321$^{c}$\hphantom{$^{abdefgi}$} & .473$^{abc}$\hphantom{$^{defgi}$} & .513$^{ab}$\hphantom{$^{cdefgi}$} & .479\\
			& $i$ & setwise.bubblesort & \textbf{.768}$^{abcdfg}$\hphantom{$^{eh}$} & .346$^{ac}$\hphantom{$^{bdefgh}$} & .388$^{bcefh}$\hphantom{$^{adg}$} & .424$^{abcfh}$\hphantom{$^{deg}$} & .754$^{abceh}$\hphantom{$^{dfg}$} & .343$^{bceh}$\hphantom{$^{adfg}$} & .479$^{abc}$\hphantom{$^{defgh}$} & .534$^{abdeh}$\hphantom{$^{cfg}$} & .505\\
			\bottomrule
		\end{tabular}
	}
	\label{tab:beir}
\end{table*}
\section{Results and Analysis}

\subsection{Effectiveness Results}\label{sec:effectiveness}

Table~\ref{tab:results} presents results for both ranking effectiveness and efficiency on TREC DL datasets. 

In regards to ranking effectiveness, it is notable that all LLM-based zero-shot ranking approaches demonstrate a significant improvement over the initial BM25 ranking. The only exception to this trend is the \textit{pointwise.qlm} approach on DL2019 across all models and DL2020 with the Flan-t5-xxl model. Interestingly, as the LLM size increases, the effectiveness of \textit{pointwise.qlm} decreases. 
This finding is particularly unexpected, given the common assumption that larger LLMs tend to be more effective. 

On the other hand, \textit{pointwise.yes\_no} method achieved a decent NDCG@10 score with Flan-t5-large when compared to other methods. However, effectiveness also did not increase as model size increased. These unexpected results for both \textit{Pointwise} methods might be attributed to the requirement of a more refined model output calibration process, ensuring their suitability for comparison and sorting across different documents~\cite{qin2023large}. 

The \textit{Listwise} approaches (\textit{listwise.generation}) are far less effective when tested with Flan-t5-large and Flan-t5-xl. However, \textit{listwise.generation} shows some improvement with Flan-t5-xxl. These results may be attributed to the fact that generating a ranking list requires fine-grained relevance preferences across multiple documents, a task that may exceed the capabilities of smaller models. In contrast, the \textit{listwise.likelihood} approach, empowered by our \textit{Setwise} prompt, markedly enhances the ranking effectiveness of the \textit{Listwise} approach, even when utilizing smaller models. We acknowledge however that \textit{listwise.likelihood} requires  access to the model output logits, whereas \textit{listwise.generation} does not. In the case of \textit{Pairwise} and \textit{Setwise} approaches, they consistently exhibit good ranking effectiveness across various model sizes and datasets.

In Table~\ref{tab:beir}, we present the zero-shot ranking effectiveness of all methods (with the exception of \textit{pairwise.allpair} due to its computationally intensive nature) across 8 widely-used BEIR datasets. 
Notably, we identify several different trends that deviate from observations made on the TREC DL datasets.

Firstly, \textit{pointwise.qlm} exhibits a slightly higher average NDCG@10 score compared to \textit{pointwise.yes\_no}. Moreover, the effectiveness of \textit{pointwise.qlm} remains stable even as the model size increases.
Secondly, \textit{listwise.generation} demonstrates comparable effectiveness to \textit{listwise.likelihood}, with the majority of gains obtained in the Touche dataset, where other methods perform worse. 
Lastly, both \textit{Pairwise} and \textit{Setwise} methods that leverage the bubble sort algorithm consistently demonstrate higher average NDCG@10 compared to when they utilize the heap sort algorithm, regardless of the model size. Overall, the variety of results we observe across different experimental settings shows the importance of not drawing conclusions about effectiveness from single datasets or model sizes.

We note that if the LLM output logits are accessible, our \textit{Setwise} approaches can also utilize these logits to estimate the likelihood of the most relevant document label. This approach eliminates the need for token generation, requiring only a single LLM forward inference to yield the output results, thus avoiding the generation of unexpected tokens.  Surprisingly, in our experiments we find that using model logits for our \textit{Setwise} approaches resulted in no change in ranking effectiveness when compare to generation, suggesting that the inference of our \textit{Setwise} approaches that fully relies on token generation is very robust.

\subsection{Efficiency Results}

\begin{figure*}
	\begin{subfigure}{1\columnwidth}
		\centering
		\includegraphics[width=\linewidth]{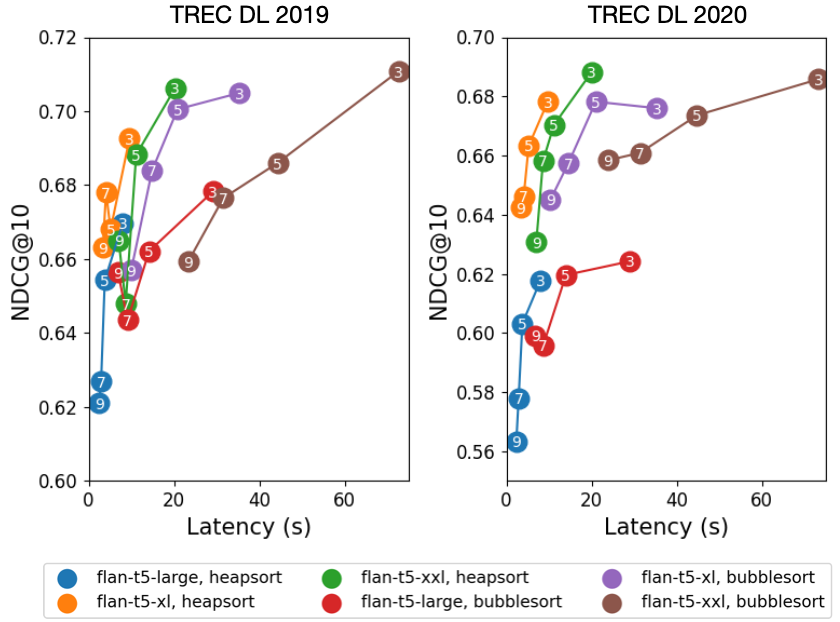}
		\caption{Setwise}
		\label{fig:trade-off:sfig1}
	\end{subfigure}
	\begin{subfigure}{1\columnwidth}
	\centering
	\includegraphics[width=\linewidth]{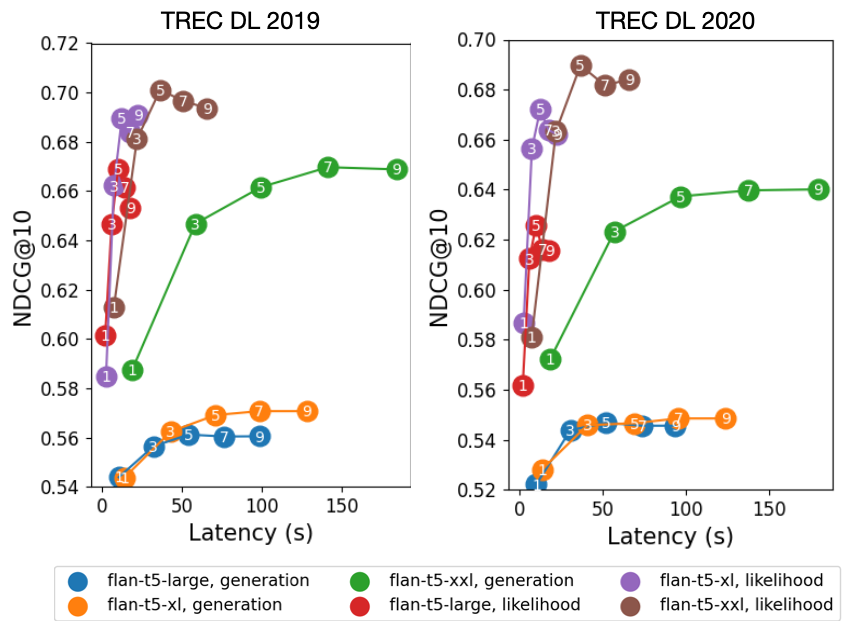}
	\caption{Listwise}
	\label{fig:trade-off:sfig2}
		\end{subfigure}
	\vspace{-6pt}
	\caption{Effectiveness and efficiency trade-offs offered by different approaches. (a -- Setwise): The numbers in the scatter plots represent the number of compared documents $c$ at each step of the sorting algorithm. (b -- Listwise) The numbers in the scatter plots represent the number of sliding windows repetitions $r$. \vspace{-10pt}}
	\label{fig:trade-off}
\end{figure*}

Regarding computational and runtime efficiency, the results presented in Table~\ref{tab:results} indicate that both \textit{Pointwise} methods exhibit fewest inference, prompt tokens, and no generated tokens. Furthermore, their computational efficiency and query latency are optimized due to efficient GPU-based batched inference. It is worth noting, however, that these methods do come with certain limitations. Specifically, they require access to the model output logits (thus currently limiting their use to just open source LLMs) and are less effective when used with larger models. In contrast, \textit{pairwise.allpair} appears to be the most expensive method that consumes the most number of prompt tokens and generated tokens due to the large number of document pair preferences needed to be inferred. Hence, even with GPU batching, \textit{pairwise.allpair} still has the worst query latency. In contrast, approaches utilizing our \textit{Setwise} prompting---namely, \textit{listwise.likelihood}, \textit{setwise.heapsort}, and \textit{setwise.bubblesort}, are far more efficient than their counterparts, \textit{listwise.generate}, \textit{pairwise.heapsort}, and \textit{pairwise.bubblesort} respectively. Notably, these improvements are achieved without compromising effectiveness. Section~\ref{sec:trade-off} will discuss further approaches on improving efficiency.

The \textit{setwise.bottlesort} and \textit{pairwise.heapsort} methods show comparable NDCG@10, but \textit{pairwise.heapsort} is cheaper. 
On the other hand, our \textit{setwise.heapsort} provides a reduction of $\approx 62\%$ in cost by only marginally reducing NDCG@10 (a 0.8\% loss).

\subsection{Effectiveness and Efficiency Trade-offs}\label{sec:trade-off}
Our \textit{Setwise} prompting is characterized by a hyperparameter $c$ controlling the number of compared documents within the prompt for each step in the sorting algorithms. In the previous experiments, we always set $c=3$. Adjusting this hyperparameter allows one to further enhance efficiency by incorporating more compared documents into the prompt, thereby reducing the number of LLM inference calls. However, we acknowledge that there is an input length limitation to LLMs (in our experiments this is 512 prompt tokens) and setting $c$ to a large value may require more aggressive document truncation, likely impacting effectiveness.

To investigate the trade-off between effectiveness and efficiency inherent in our \textit{Setwise} approach, we set $c = 3, 5, 7, 9$ while truncating the documents in the prompt to $128, 85, 60, 45$ tokens, respectively. This reduction in document length is necessary to ensure prompt size is not exceeded. The NDCG@10, along with query latency for all models while varying $c$, is visualized in Figure~\ref{fig:trade-off}a for the TREC DL datasets. As expected, larger $c$ reduces query latency but often degrades effectiveness. Notably, the heap sort algorithm consistently proves more efficient than bubble sort. For instance, with Flan-t5-xl and $c=9$, heap sort achieves strong NDCG@10 with a query latency of $\approx$3 seconds. When compared to the other methods outlined in Table~\ref{tab:results}, this represents the lowest query latency, except for the \textit{Pointwise} approaches with Flan-t5-large, albeit with superior ranking effectiveness. 
It's worth noting that the ranking effectiveness decline with larger $c$ values could also be attributed to the increased truncation of passages. LLMs with extended input length capacity might potentially yield improved ranking effectiveness for larger $c$. This area warrants further exploration in future studies.

Similarly, the \textit{Listwise} balance effectiveness and efficiency through the adjustment of the repetition count $r$ for the sliding window. In our prior experiments, we consistently set $r=5$ to ensure that at least 10 of the most relevant documents can be brought to the top. In Figure~\ref{fig:trade-off:sfig2}, we investigate the influence of varying $r$ on \textit{Listwise} approaches. Latency exhibits a linear relationship with $r$, which aligns with expectations. A larger value of $r$ can enhance the effectiveness of \textit{listwise.generate}, and beyond $r>5$, the improvement levels off. Conversely, the \textit{listwise.likelihood} approach, which leverages our \textit{Setwise} prompting, showcases notably higher effectiveness and efficiency. Even with a small value of $r$ the performance of \textit{listwise.likelihood} exceeds that of \textit{listwise.generate}, with the highest performance achieved around $r=5$.

\begin{figure*}
	\begin{subfigure}{1.6\columnwidth}
		\centering
		\includegraphics[width=\linewidth]{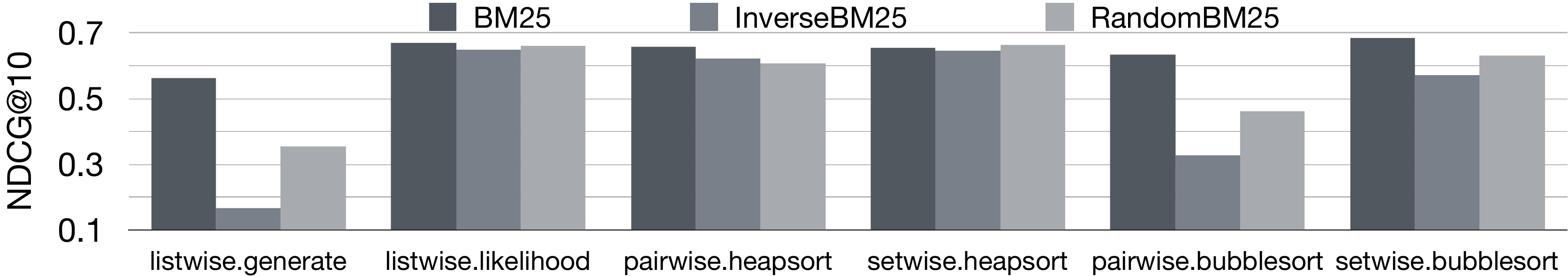}
		\caption{TREC DL 2019}
		\vspace{8pt}
		\label{fig:init_ranking:sfig1}
	\end{subfigure}
	\begin{subfigure}{1.6\columnwidth}
		\centering
		\includegraphics[width=\linewidth]{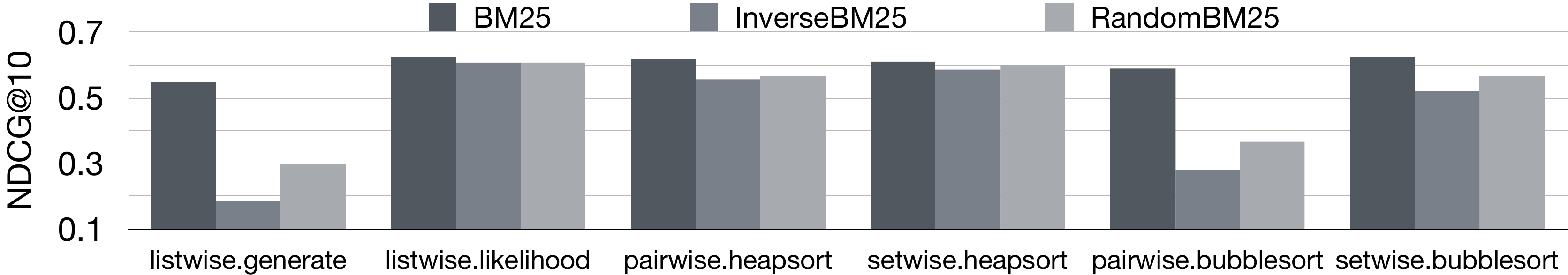}
		\caption{TREC DL 2020}
		\label{fig:init_ranking:sfig2}
	\end{subfigure}
	\caption{Sensitivity to the initial ranking. We use Flan-t5-large and $c=4$ for the Setwise approach. }
	\label{fig:init_ranking}
\end{figure*}

\vspace{-6pt}
\subsection{Sensitivity to the Initial Ranking}
The ranking effectiveness of the original \textit{Listwise} and \textit{Pairwise} methods is influenced by the initial ranking order~\cite{sun2023chatgpt,qin2023large}. 
To investigate this aspect in relation to our approach, we consider different orderings of the initial BM25 list; specifically, 1) initial BM25 ranking; 2) inverted BM25 ranking; and 3) random shuffled BM25 ranking. Each of these initial rankings was used to test different reranking methods using Flan-t5-large. The results are presented in Figure~\ref{fig:init_ranking}.
Different initial ranking orders negatively impact \textit{listwise.generate}, \textit{pairwise.heapsort} and \textit{pairwise.bubblesort}; \textit{pairwise.heapsort} is the most robust method. These findings align with the literature~\cite{sun2023chatgpt,qin2023large}. 

In contrast,  \textit{Setwise} prompting  is far more robust to variations in the initial ranking order. Both \textit{listwise.likelihood} and \textit{setwise.bubblesort} exhibit large improvements over \textit{listwise.generate} and \textit{pairwise.bubblesort}, in the case of the inverted BM25 ranking and randomly shuffled BM25 ranking. Moreover, they demonstrate a similar level of robustness to \textit{pairwise.heapsort}. This leads us to the conclusion that our \textit{Setwise} prompting approach substantially enhances the zero-shot re-ranking with LLMs in relation to the initial ranking.

\vspace{-4pt}
\subsection{Effectiveness and Costs of other LLMs}
In the previous sections, we only used Flan-T5 as the backbone LLM. Flan-T5 is a transformer encoder-decoder model.
In this section, to better understand the impact of different models, we investigate popular transformer decoder-only LLMs (open-sourced: llama2-chat-7b\footnote{\url{https://huggingface.co/meta-llama/Llama-2-7b-chat-hf}}~\cite{Touvron2023Llama2O}, vicuna-13b-v1.5\footnote{\url{https://huggingface.co/lmsys/vicuna-13b-v1.5}}~\cite{zheng2023judging}; closed-source: OpenAI gpt-3.5-turbo-1106\footnote{\url{https://platform.openai.com/docs/models/gpt-3-5}}) on DL2019 and DL2020. 
For open-source models we measure cost  in terms of query latency in seconds (s); for closed models we measure  the API call costs in US dollars (\$).\footnote{We exclude \textit{Pointwise} approaches as they are less effective, and \textit{pairwise.bubblesort} for gpt-3.5-turbo as it is too costly.}
Results are presented in the Table~\ref{tab:results_llms}.

	Flan-T5 in our previous results is a better backbone than Llama2 and Vicuna, regardless of ranking method.
	Notably, \textit{Setwise} exhibits the best overall performance when considering these models, showcasing its robustness.
	 For gpt-3.5-turbo (closed model), we compared \textit{Setwise} and \textit{Listwise} using 10 documents at the time ($c=10$), as this LLM has a longer input context limit; for \textit{Listwise}, we set window size 10, step size of 5 and repeat sorting twice for fair comparison with \textit{Setwise}. \textit{Listwise} achieves the highest effectiveness; however, \textit{Setwise} achieves similar effectiveness (no significant differences) but at only half the cost. 
	 
	 \begin{table}[t!]
	 	\centering
	 	\caption{
	 		Results obtained with other LLMs on TREC DL datasets: We report the cost in terms of query latency in seconds (s) if the LLM's weights are publicly available, or the API call costs in US dollars (\$) if the LLM is only accessible via API calls.
	 	}
	 	\resizebox{1\columnwidth}{!}{
	 		\begin{tabular}{cc|l|ll|ll}
	 			\toprule
	 			\multicolumn{3}{c}{} & \multicolumn{2}{|c|}{\textbf{TREC DL 2019}} &  \multicolumn{2}{|c}{\textbf{TREC DL 2020}}
	 			\\
	 			\toprule
	 			& {\small \textbf{\#}}
	 			&{\small \textbf{Methods}}
	 			& {\small \textbf{NDCG@10}}
	 			& {\small \textbf{Cost(s or \$)}}
	 			& {\small \textbf{NDCG@10}}
	 			& {\small \textbf{Cost(s or \$ )}} \\ 
	 			\midrule
	 			\multirow{5}{*}{ \rotatebox[origin=c]{90}{llama2-chat-7b}}
	 			& $a$ &
	 			listwise.generation &
	 			.508\hphantom{$^{bcde}$} & 144.9 s & .475$^{c}$\hphantom{$^{bde}$} &  143.9 s\\
	 			&$ b$ &
	 			pairwise.bubblesort & 
	 			.538$^{ac}$\hphantom{$^{de}$} & 64.3 s&.503$^{ac}$\hphantom{$^{de}$}& 61.5 s\\
	 			&$ c$ &
	 			pairwise.heapsort &
	 			.465\hphantom{$^{abde}$} & 18.3 s &.416\hphantom{$^{abde}$} &  17.9 s\\
	 			& $d$ &
	 			setwise.bubblesort &
	 			\textbf{.578}$^{abc}$\hphantom{$^{e}$} & 38.2 s& \textbf{.530}$^{ac}$\hphantom{$^{be}$}&  38.5 s\\
	 			& $e$ &
	 			setwise.heapsort &
	 			.568$^{c}$\hphantom{$^{abd}$} &\textbf{ 10.9} s& .514$^{c}$\hphantom{$^{abd}$} & \textbf{10.9} s \\
	 			\bottomrule
	 			
	 			\multirow{5}{*}{ \rotatebox[origin=c]{90}{vicuna-13b}}
	 			& $a$ &
	 			listwise.generation &
	 			.639\hphantom{$^{abce}$}   & 225.9 s& \textbf{.608}\hphantom{$^{bcde}$} &  226.3 s\\
	 			& $b$ &
	 			pairwise.bubblesort & 
	 			.612\hphantom{$^{abce}$}  &166.6 s & .592\hphantom{$^{acde}$}&  169.0 s\\
	 			& $c$ &
	 			pairwise.heapsort &
	 			.617\hphantom{$^{abce}$}  & 46.0 s & .582\hphantom{$^{abde}$} & 45.6 s\\
	 			& $d$ &
	 			setwise.bubblesort &
	 			.622\hphantom{$^{abce}$}  &57.6 s & .602\hphantom{$^{abce}$} & 58.8 s \\
	 			& $e$ &
	 			setwise.heapsort &
	 			\textbf{.659}$^{bcd}$\hphantom{$^{a}$}& \textbf{16.7} s& .583\hphantom{$^{abcd}$} & \textbf{16.4} s \\
	 			\bottomrule
	 			\multirow{4}{*}{ \rotatebox[origin=c]{90}{gpt-3.5}}
	 			& $a$ &
	 			listwise.generation &
	 			\textbf{.712}\hphantom{$^{bcd}$} & 0.045 \$& \textbf{67.2}\hphantom{$^{bcd}$} & 0.046 \$\\
	 			& $b$ &
	 			pairwise.heapsort & 
	 			.694\hphantom{$^{acd}$} & 	0.171 \$&65.1\hphantom{$^{acd}$}& 0.169 \$	\\
	 			& $c$ &
	 			setwise.bubblesort &
	 			.699\hphantom{$^{abd}$}  & 0.084 \$&65.9\hphantom{$^{abd}$} & 	0.088  \$	\\
	 			& $d$ &
	 			setwise.heapsort &
	 			.693\hphantom{$^{abc}$} & \textbf{0.029}  \$& 65.6\hphantom{$^{abc}$} &  \textbf{0.028} \$\\
	 			\bottomrule
	 		\end{tabular}
	 	}
	 	\label{tab:results_llms}
	 \end{table}
	 
	 We note that there could be potential data contamination with instruction-tuned LLMs: during the instruction fine-tuning tasks, these models could have been fine-tuned on the MS MARCO or BEIR datasets. Although the document ranking task and the ranking prompts used in this paper are unlikely to be part of the instruction fine-tuning dataset used for these LLMs, we acknowledge that data contamination could still artificially impact the effectiveness of the considered LLMs in ranking tasks. However, since we conducted experiments with different methods based on the same instruction-tuned model, we believe our empirical comparison still offers insights for the practical use of LLM-based re-rankers.

\section{Conclusion}

We undertook a comprehensive study of existing LLM-based zero-shot document ranking methods, employing strict and consistent experimental conditions. Our primary emphasis was on evaluating both their ranking effectiveness and their efficiency in terms of computational efficiency and runtime latency --- factors that are often disregarded in previous studies. Our findings unveil some unforeseen insights, and effectiveness-efficiency trade-offs between different methods. This information equips practitioners with valuable guidance when selecting the most appropriate method for their specific applications. 

To further boost efficiency of LLM-based zero-shot document ranking, we introduced an innovative \textit{Setwise} prompting strategy. \textit{Setwise} has the potential to enhance both effectiveness and efficiency for \textit{Listwise} approaches provided the model logits are accessible. \textit{Setwise} also notably enhances the efficiency of sorting-based \textit{Pairwise} approaches. Furthermore, \textit{Setwise} prompting offers a straightforward way to balance effectiveness and efficiency by incorporating more documents for comparison in the prompt. Additionally, approaches equipped with  \textit{Setwise} prompting demonstrated strong robustness to variation in the initial retrieval set used for reranking.

Future work should focus on evaluating the \textit{Setwise} prompting approach on a wider array of LLMs, including LLaMA models~\cite{touvron2023llama,touvron2023llama2} as well as the OpenAI LLM APIs. 
Additionally, recent advanced self-supervised prompt learning techniques~\cite{yang2023large,fernando2023promptbreeder} could be used to refine the \textit{Setwise} approach.


\balance
\bibliographystyle{ACM-Reference-Format}
\bibliography{llmranker}

\appendix

\end{document}